\documentclass{desyproc}
\newcommand{\mr}{\mathrm}
\newcommand{\bb}{b$\bar{\mr b}$}
\newcommand{\hbb}{H~$\rightarrow$~\bb}

\begin{document}
%------------------------------------
\title{Measuring Central Exclusive Processes at LHC}

%for single authors the superscripts are optional
\author{{\slshape Marek Ta\v{s}evsk\'{y}}\footnote{Talk given at the 13th
International Conference on Elastic and Diffractive Scattering, July 2009, 
CERN}\\[1ex]
Institute of Physics, Academy of Sciences of the Czech republic, 
Na Slovance 2, 182 21 Prague, Czech republic}

% if the proceedings are available online (e.g. at Indico)
% please enter the contribution ID or file_name below for the DOI
%\contribID{32}
%\contribID{smith\_joe}

% TO THE CONFERENCE EDITORS: 
% please update the following information      
% before sending the template to the authors
% \confID{800}  % if the conference is on Indico uncomment this line
\desyproc{DRAY-PROC-2009-xx}
\acronym{EDS'09} % if you want the Acronym in the page footer uncomment this 
%line
%\doi  % if there is an online version we will register DOIs

\maketitle

\begin{abstract}
Diffractive physics program for ATLAS and CMS is discussed with emphasis on 
measurements of central exclusive processes. At low luminosities, a L1 trigger
based on requiring rapidity gaps can be used, while at high luminosities, the 
use of proton taggers proposed to be placed at 220~m and 420~m from the 
interaction point is foreseen. 
\end{abstract}

\section{Introduction}\label{Intro}
%Diffractive physics program includes a variety of processes whose cross 
%sections span several orders of magnitude. In the following, they are thus 
%divided into two categories, as processes measurable at low luminosities where
%there is no pile-up background and at high luminosities where the pile-up is 
%the most dangerous source of background. 
The central exclusive production (CEP)
of new particles has received a great deal of attention in recent years (see
\cite{FP420TDR} and references therein). 
The process is defined as $pp\rightarrow p\oplus\phi\oplus p$
and all of the energy lost by the protons during the interaction
(a few percent) is used in the production of the central system, $\phi$. The 
final state therefore consists of a centrally produced hard subprocess, two 
very forward protons and no other activity. The '$\oplus$' sign denotes the
regions devoid of activity, often called rapidity gaps. In Double Pomeron 
Exchange (DPE), the central system contains remnants from the diffractive
exchange in addition to the hard subprocess. More details on the 
diffraction physics program of the ATLAS experiment can be found e.g. in 
\cite{MT-photons}.

\section{Low luminosity running}\label{Lowlumi}
At low luminosity, the diffractive processes can
be detected using rapidity gaps. A possible L1 trigger would be based on a
requirement of a rapidity gap on one or both sides from the IP and an activity
in the central detector with energy over a certain threshold. 
%($\sim$ 20--30~GeV). 
The gap may span the region from the forward calorimeters of the ATLAS 
detector \cite{ATLASTDR} or CMS detector \cite{CMSTDR} over the luminosity 
detectors (LUCID \cite{LUCID} in ATLAS or TOTEM \cite{TOTEM} 
in CMS) up to ZDC detectors in ATLAS \cite{ZDCATLAS} or in CMS \cite{CMSTDR}.  
Measurements which would be 
straightforward and hence suitable for analyses of the very early LHC data are
ratios of the kind of X$+$gaps/X(incl.), where X may be W, Z, dijet, heavy 
quark and dilepton, and X(incl.) means measuring X without requiring rapidity 
gaps. Measurements of ratios are convenient since many sources of systematic 
uncertainties are canceled, particularly that of the luminosity at the early
phase and among other, they also serve as valuable 
checks of different components of the formalism used to predict the CEP 
cross section by the KMR group \cite{KMR,KMR-08}. The soft survival 
probability, $S^2$, can be studied in electroweak processes, such as W$+$gaps 
or Z$+$ gaps. $S^2$ is defined as a probability that 
additional soft secondaries will not populate the gaps and it explains the
factorization breaking observed at hadron colliders when diffractive parton
density functions (dPDF) obtained in Single Diffraction (SD) at HERA were 
applied in measurements of SD by CDF \cite{CDF-factbreak}. 
%$S^2$ can also be obtained from a ratio of dPDFs measured in SD and DPE 
%processes. 
The generalized gluon distribution, 
$f_g$, can be probed in exclusive $\Upsilon$ production proceeding via either 
a photon or an odderon exchange. The higher-order QCD effects,
especially Sudakov-like factors and also a possible role of the enhanced 
absorptive corrections can be studied in exclusive two- or three-jet events. 
When the proton tagging becomes available, the t-dependence of the elastic, 
SD and DPE cross sections can be obtained and hence effect of individual 
components of the pile-up background can be evaluated.

\subsection{Dijet production in DPE and CEP}
Without proton tagging, the dijet production in DPE and CEP can be measured
by requiring two central jets and rapidity gaps on both sides of the IP in
forward calorimeter, LUCID/TOTEM and ZDC. In DPE, the rapidity gaps may be
spoiled by particles from the pomeron remnants and although the cross section
is about two orders of magnitude larger than the CEP cross section at the same 
dijet mass, the CEP cross section will dominate if the forward calorimeter is
required to be devoid of activity. The measurement of the
dijet production in CEP at 7--14~TeV may be compared with a similar 
measurement made at Tevatron from which models used to describe the data may 
be constrained. 
%The exclusive dijet 
%production is then special in a sense that by restricting the phase space to 
%certain regions, we may be able, in one process, to separate different 
%effects on which the CEP prediction is based. \cite{KMR-08}.

\subsection{Gaps between jets}
%Another interesting process is the dijet production via color singlet exchange
%for which we require two jets, each in opposite forward calorimeter and a 
%rapidity gap in the central detector. Compared to the existing measurement
%by D0 at cms energy of 1.8~TeV \cite{D0-cse}, ATLAS can make an improvement 
%thanks to the increased cms energy and hence the available phase space. 
By selecting events with two jets each in opposite side of forward calorimeter
and a rapidity gap in the central detector, ATLAS and CMS can improve an 
existing 
measurement of this type by D0 \cite{D0-cse} at centre-of-mass (c.m.s.) energy
of 1.8~TeV. Different colour singlet exchange
models can be tested by comparing data with predictions for the gap fractions
as functions of rapidity between the jets.   

%Exclusive dimuon production in gammagamma collisions tagged with forward 
%rapgaps 

\section{High luminosity running}
A great attention is recently devoted to the possibility of
complementing the standard LHC physics menu by adding forward proton detectors
(FPD) to the ATLAS and CMS detectors. They would
detect a great part of the energy flow that escapes undetected by the main 
detectors.
%A well-known feature of hadronic production at the LHC is that the particle 
%multiplicity is peaked in the central region, while most of the energy flows 
%very forward and undetected by the main detector. In order to have a chance 
%to explore the very rich forward physics it is necessary to instrument the 
%forward region of the main detectors.

\subsection{SM and BSM Higgs boson production}
\label{physics}
The forward proton tagging will provide an exceptionally clean
environment to search for new phenomena at the LHC and to identify their
nature. Of particular interest in this context is the CEP which gives
access to the generalized (or skewed) PDFs.
%At the highest available luminosities, CEP may become a discovery channel
%for particles with appropriate quantum numbers that couple to gluons. 
The CEP of a SM (or MSSM) Higgs boson is attractive for two reasons: firstly, 
if the outgoing
protons remain intact and scatter through small angles then, to a good
approximation, the central system $\phi$ must be produced in a $J_z=0$, CP even
state, therefore allowing a clean determination of the quantum numbers of any
observed resonance. Here $J_z$ is the projection of the total angular momentum 
along the proton beam axis. 
Secondly, from precise measurements of proton momentum
losses, $\xi_1$ and $\xi_2$, the mass of the central system can be measured
much more precisely than from the dijet mass measured in the calorimeters, by 
the so-called missing mass
method, $M^2=\xi_1\xi_2 s$, which is independent of the decay mode.
The simplest decay mode from an experimental perspective is the WW decay mode,
in which one (or both) of the W bosons decay leptonically. With standard single
and double lepton trigger thresholds at ATLAS (or CMS), approximately 6 events
are expected for Higgs boson mass around 160 GeV with luminosity of
30~fb$^{-1}$ \cite{WWpaper}. In the \bb\ decay mode, the quantum number
selection rules in CEP
strongly suppress the QCD b-jet background, nevertheless severe requirements
necessary to get rid of the pile-up background make the event yield rather
modest \cite{CMS-Totem,SMHbb-ATLAS}. Full details of the calculation of the
background to this channel are described in \cite{KMRDO,HKRSTW}.

\begin{figure}[h]
\hspace*{1.5cm}\includegraphics[width=.4\textwidth,height=3.5cm]{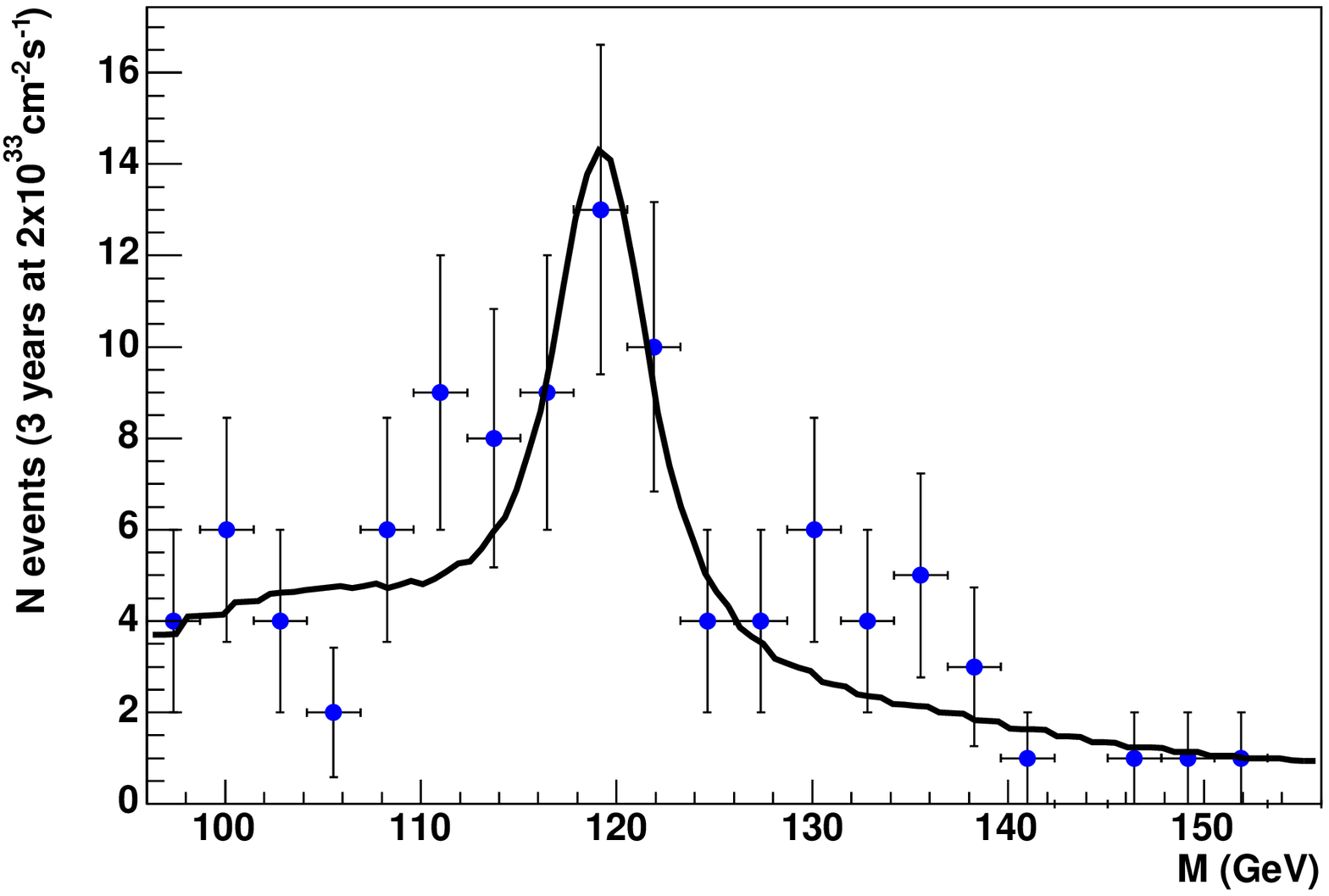}
\includegraphics[width=.4\textwidth,height=3.5cm]{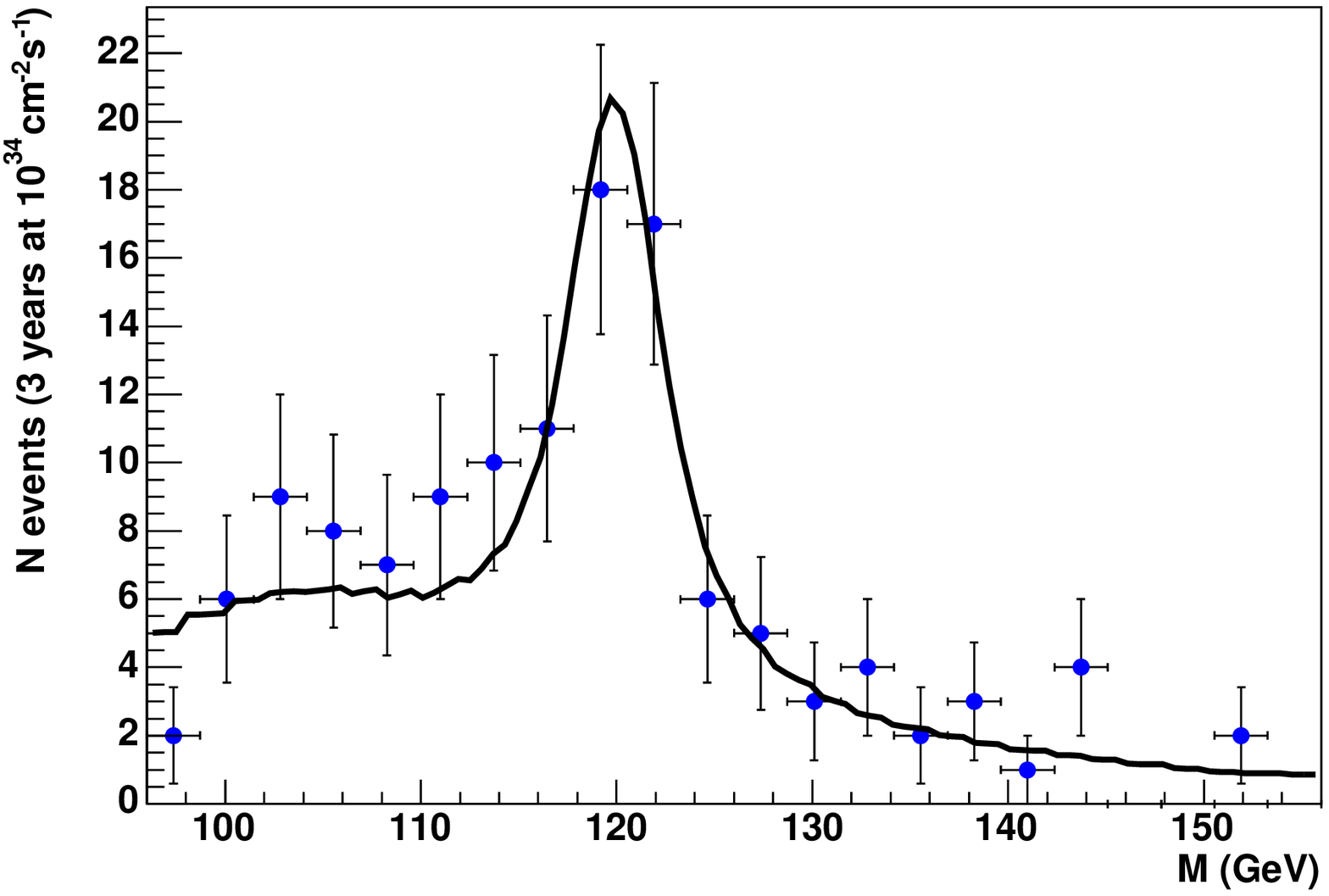}
\caption{A typical mass fit of the $H\rightarrow$ \bb\ signal
and its backgrounds for 3 years of data taking with ATLAS and the 420$+$420 
detector configuration (a) at $2\cdot 10^{33}$cm$^{-2}$s$^{-1}$ (60 fb$^{-1}$).
The significance of the fit is $3.5\sigma$. (b) at 
$10^{34}$cm$^{-2}$s$^{-1}$ after removing the pile-up background completely. 
The significance is $5\sigma$. Figures from \cite{CLP}.}
\label{mass_spectrum}
\end{figure}

In certain regions of the MSSM parameter space the cross section for
the CEP of Higgs bosons is significantly enhanced and hence
making the \bb\ decay mode attractive \cite{HKRSTW,CLP}. 
In Fig.~\ref{mass_spectrum} an example mass spectrum is shown 
for MSSM Higgs boson candidates of mass of 120~GeV decaying into \bb\ for 
$\tan\beta=40$ 
(corresponding to the final cross section of about 18~fb) after 
3~years of data taking
at luminosity of $2\cdot 10^{33}$ cm$^{-2}$s$^{-1}$ 
or 3 years at $10^{34}$ cm$^{-2}$s$^{-1}$. At the low luminosity,
the pile-up background can be completely eliminated and the statistical 
significance is around 3.5$\sigma$. At the highest luminosity, fast timing
detectors are necessary to reduce the pile-up background - significance of
5$\sigma$ is achieved with time resolution of 2~ps (see Section \ref{timing}).
%Another interesting feature
%coming from the MSSM studies is that the Higgs boson mass spectrum gets 
%broader with increasing $\tan\beta$ which from a certain value of $\tan\beta$
%may serve as a distinguishing criterion between the SM and MSSM signals 
%\cite{HKRSTW}.
\\
Encouriging results have recently been obtained for the CEP Higgs bosons 
in the \bb\ and $\tau\tau$ decay channels in the model which extends the SM 
by a fourth generation of heavy fermions \cite{4gen,Sven-eds09}. 
NMSSM provides the $h\rightarrow aa \rightarrow \tau\tau\tau\tau$ channel 
which in CEP gives a very good signal-to-background ratio and is interesting 
from the experimental perspective \cite{NMSSM}. Also, an excellent mass 
measurement of the neutral Higgs boson decaying into \bb\ in the triplet model
is possible in the forward proton mode \cite{triplet}.
 
\subsection{Central exclusive jet production}
The tagging of both protons in FPDs will enable a measurement of
the proton transverse momenta and azimuthal angles which allows us to study
the opacity of the incoming protons, and more generally, to test the dynamics
of the soft survival probability by studying the correlations between the 
outgoing protons \cite{KMR,CDF-O}. This can be carried out with the CEP of 
dijets as the cross section is large. Thanks to the $J_z = 0$ selection rule 
which
is applicable to all CEP processes, quark jet production is suppressed and the 
CEP can then be recognized as reduced ratio of b-jets to all jets when 
compared to other production processes.

\subsection{Diffraction and QCD}
%At low luminosity, SD production of mesons, dijet and vector bosons can be 
%studied. At higher luminosities, DPE can be used for 
%similar studies, the lower event rate being compensated by additional 
%rejection against the combinatorial pile-up backgrounds (see 
%Section~\ref{timing}). 
The SD and DPE processes serve to provide an information about 
the low-x structure of the proton and the dPDFs. 
Inclusive jet and heavy quark production are mainly sensitive to
the gluon component of the dPDFs, while vector boson production is sensitive 
to quarks. The kinematic region covered expands that explored at HERA and 
Tevatron, with values of $\beta$ 
(the fractional momentum of the struck parton in the diffractive exchange )
as low as $10^{-4}$ and of Q$^2$ up to $10^4$ GeV$^2$. More information about
the SD dijet and W boson production can be found elsewhere in these 
proceedings \cite{Vojta,Carvalho}.

%The cross section for dijet production via DPE is expected to be 
%\cite{CMS-Totem} approximately 45~pb for $E_T>$ 50~GeV (after applying 
%$S^2$ = 0.03). This process can be investigated at high luminosity, although 
%the number of observed events will depend upon the jet threshold pre-scale at 
%L1. Dijet production in SD has an expected cross section 
%\cite{CMS-Totem} of 1.5~nb for $E_T>$ 50~GeV (after applying $S^2$ = 0.05 
%\cite{KMR-06}). 
%An expected number of dijet events produced in the ATLAS 
%detector in a single day at luminosity of $10^{32}$cm$^{-2}$s$^{-1}$ is more 
%than 10000 for SD and several hundred for DPE. 

%\subsubsection{Vector boson production}
%The W and Z boson production in SD and DPE events can also be studied with the
%ATLAS detector. So far, the studies have been performed by CMS. More details
%can be found elsewhere in these proceedings, see \cite{Monika}. 
%SD and DPE production of B-mesons
%with $B\rightarrow J/\psi X$ and $J/\psi\rightarrow\mu^+\mu^-$ was studied for
%CMS and more information can be found in \cite{CMS-Totem}.    

\subsection{Photon-photon physics}
As the LHC beams act also as a source of high-energy photons a rich program of
photon-photon and photon-proton physics can be pursued. Photon-induced 
processes have been extensively studied at LEP and HERA. However at LHC, these
processes can be investigated in an unexplored region of the phase-space. The 
final state topology
is similar to CEP, i.e. a central system, X, separated on each side by large 
rapidity gap from a very forward proton detected in the FPD. Different average
proton transverse momenta make it possible to separate between diffractive 
and photon-induced events.

The W- and Z-pair production (as a tool to study anomalous triple and quartic 
gauge couplings) is discussed elsewhere in these proceedings \cite{Olda}. The
SUSY particle production is described in \cite{Schul}. 

\subsubsection{Lepton pair production}
Two-photon exclusive production of muon pairs has a well known QED cross 
section, including very small hadronic corrections \cite{KMRO}. 
Very recently, such event candidates have been observed by the CDF 
\cite{CDF-excllept,Albrow} and their cross sections found in a good agreement 
with theory. After applying simple selection criteria such as $p_T>$ 10~GeV, 
$|\eta | <$ 2.5 and requiring one forward proton tag, the cross section is 
1.3~pb \cite{FP420TDR,Louvain-photon}. This corresponds to approximately 
50~muon pairs detected in a 12~hour run at a mean luminosity of 
$10^{33}$cm$^{-2}$s$^{-1}$. The large event rate coupled with a 
small theoretical uncertainty makes this process a perfect candidate for the 
absolute LHC luminosity calibration \cite{KrzP} and also of the FPD system at 
420~m \cite{FP420TDR}. 
The $e^+e^-$ production can also be studied, although the 
trigger thresholds will be larger and hence the final event rate reduced.

\subsection{Photoproduction}
The high luminosity and the high c.m.s. energies available for photoproduction 
processes at the LHC allows us to study electroweak interactions and to search
Beyond the Standard Model up to the TeV scale \cite{Louvain-photon}. 
%There is a number of processes with sizable cross sections that can be 
%studied at low LHC luminosities. 

\subsubsection{Associated WH production}
As shown in \cite{Louvain-photon}, the cross section for
the associated WH production ($pp \rightarrow (\gamma p \rightarrow 
WHq^{\prime}) \rightarrow pWHq^{\prime}Y$ after applying selection criteria
and considering five different final states is 0.17~fb at m$_H$ = 115~GeV and
0.29~fb at m$_H$ = 170~GeV. 
%reaches 23~fb at m$_H$ = 115~GeV and 17.5~fb at m$_H$ = 170~GeV. Considering 
%five different final states and only irreducible backgrounds, the ratios of 
%signal to background cross sections after applying all selection cuts are 
%0.17~fb/42.1~fb at m$_H$ = 115~GeV and 0.29~fb/1.72~fb at m$_H$ = 170~GeV. 
The most promising channel seems to be  the $jjl^{\pm}l^{\pm}$ at 
m$_H$ = 170~GeV where the signal to irreducible background ratio is 
0.22~fb/0.28~fb, so luminosity of 100 fb$^{-1}$ might reveal the $HWW$ gauge 
coupling.

\subsubsection{Single top quark and anomalous top quark production}
Photoproduction of single top quark ($pp \rightarrow (\gamma p \rightarrow 
Wt) \rightarrow pWtY$) is dominated by t-channel amplitudes in
association with a W boson which all are proportional to the CKM matrix 
element $|V_{tb}|$. The ratio of associated Wt production cross section
to the sum of all top production cross sections is 5\% for parton-parton 
interactions, while it is 50\% in photoproduction. In \cite{Louvain-photon} two
topologies were studied, namely $lbjj$ and $llb$. The signal cross section
after selection cuts of about 44~fb with a signal to irreducible background
ratio of 0.6 suggest that this mechanism and hence $|V_{tb}|$ may be 
easily measurable even with luminosity of 1~fb$^{-1}$.

At LHC the exclusive single top quark photoproduction can only occur via 
flavour changing neutral current processes which are not present at 
tree level of SM but appear in many extensions of SM such as two Higgs-doublet
models or R-parity violating supersymmetry. The final state of this $pp 
\rightarrow (\gamma p \rightarrow t) \rightarrow ptY$ process is composed of a
b-jet and a W boson. In \cite{Louvain-photon} the leptonic $lb$ topology was 
studied and only photoproduction $\gamma p \rightarrow W+jet$ background 
considered. With an integrated luminosity of 1~fb$^{-1}$ the expected 
limits for anomalous couplings $k_{tu\gamma}$ and $k_{tc\gamma}$ at 95\% CL 
are greatly improved with respect to existing best estimates. 
%the limits on 
%the $k_{tc\gamma}$ coupling may be obtained for the first time at all. 

\subsubsection{Photoproduction of jets}%\vspace*{-0.1cm}
In photoproduction of jets, the fraction of the photon, $x_{\gamma}$, and 
proton, $x_p$, 
four-momentum carried by a parton involved in binary hard scattering is 
calculated using the energies and angles of jets in the central detector and 
of the protons in the FPDs. The direct photon processes
are characterized by $x_{\gamma}\sim 1$ and resolved photon processes by 
$x_{\gamma}<1$. The H1 and ZEUS collaborations have constrained the region 
$x_p, x_{\gamma}>0.1$ for diffractive photoproduction. At LHC
%due to higher 
%proton beam energy and available photon energies and due to a larger rapidity 
%coverage of the main ATLAS detector and a slightly higher $E_T$ cut of
%jets ($\sim$20~GeV) compared to HERA conditions ($\sim$5~GeV), 
we expect 
to reach values 
of $x_p$ and $x_{\gamma}$ of an order of magnitude lower than at HERA. 
Furthermore, the diffractive photoproduction of dijet systems at the LHC 
promises to shed light on the issue of the QCD factorization breaking recently 
reported in the same process by the H1 experiment \cite{factbreak}.
%H1 observe that the the ratio of the measured di-jet cross section to the NLO 
%prediction, based on diffractive parton densities from DIS data, is a factor 
%0.5 ESCN1 0.1 smaller in photoproduction than the same ratio in DIS. This 
%suppression is a clear observation of QCD hard scattering factorization 
%breaking at HERA.

\subsubsection{Exclusive $\Upsilon$ production}
Exclusive $\Upsilon$ photoproduction, $\gamma p \rightarrow \Upsilon p$ can be 
studied using FPDs \cite{KMR-08}, although only one proton 
can be tagged due to the low mass of the $\Upsilon$. The cross section is 
expected to be approximately 1.25~pb for the decay channel 
$\Upsilon\rightarrow\mu^+\mu^-$ and is sensitive to the same skewed 
unintegrated gluon densities of the proton as the CEP of Higgs boson.
Measuring this process 
thus helps to constrain the $f_g$ as the soft survival factor is expected to 
be close to 1. The $\gamma p \rightarrow \Upsilon p$ process can also occur 
via odderon exchange and this channel could be the first evidence for the 
odderon's existence. 
%This process is also expected to be of help to calibrate and align the FPDs.

\section{Proposed forward proton taggers for high luminosities}
%In both ATLAS and CMS, the proposals are
%based on the common R\&D work by the FP420 collaboration \cite{FP420TDR}. 
%In ATLAS in addition, the region of 220~m is also proposed to be equipped by 
%the same forward detectors as at 420~m and a common project, 
%the ATLAS Forward Proton (AFP), was formed to pursue this goal. 
\subsection{FP420}\label{FP420}
The FP420 R\&D collaboration \cite{FP420TDR}, with members from ATLAS, CMS and 
LHC studied the possibility of installing high precision tracking and timing 
detectors at 420~m from the IP. Detection of the protons will be achieved by 
two 3D silicon detector stations at each end of the FP420 region. This novel
technique provides high radiation-resistive detectors close to the beam with 
an insensitive area as small as 5~$\mu$m and with a resolution of about 
15~$\mu$m. The tracking and timing detectors will be attached to a movable
beam pipe. As the beam pipes in the 420~m region are contained in an 
interconnecting cryostat and the sensitive detectors are best operated at room
temperature, a new connection cryostat has been designed using a modified Arc 
Termination Module at each end.
%a new connection cryostat with approximately 8~m of room temperature beam 
%pipes has been designed using a modified Arc Termination Module (which 
%includes cold to warm transitions for the beam-pipes) at each end.

\subsection{AFP}
The AFP project (ATLAS Forward Proton) \cite{AFPLOI} is a combined effort to 
install FPDs
at 220 and 420~m from the IP of ATLAS. The 420~m region has been described 
above and the proposed equipment for the 220~m region is very similar; 
differences are mainly in no need to change the cryostat and an addition of a
detector to be used for L1 trigger.  
The detectors will be read out by standard ABCNext chip being developed for the
silicon detectors in ATLAS. 

\subsection{Acceptance and Resolution}\label{acceptance}
%In general, the position of a proton hit in detectors at 220 or 420 m depends
%(for a given beam optics) on the energy and the scattering angle of the
%outgoing proton and the z-vertex position of the collision. The energy and
%scattering angle are directly related to the kinematic variables $\xi$ and
%$-t$. 
With the position resolution of 15~$\mu$m we expect a mass resolution of the 
order of 1--2\% for the 420+420 and about 3\% for the
420+220 configurations over a mass range of 120--200~GeV. For given dipole 
apertures and collimator settings and a 
thin window of 200 $\mu$m, the expected $\xi$ range is 0.002--0.02 for
420+420 and 0.01--0.15 for the 420+220 configuration. 

\begin{figure}[h]
\centerline
{\includegraphics[width=.5\textwidth,height=4cm]{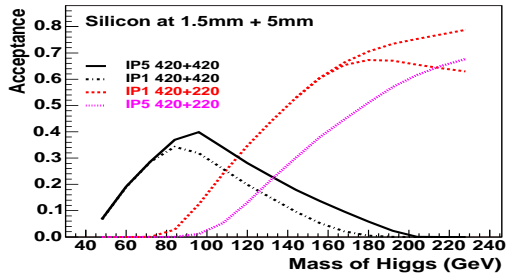}}
\caption{Mass acceptance for the 420+420 and 420+220 detector configurations 
(5~mm from the beam for 420~m and 1.5~mm for 220~m detectors) for IP1 and IP5.
From \cite{FP420TDR}.}
\label{mh-acc}
\end{figure}

The low-$\xi$ (and therefore low mass) acceptance depends critically on the
distance of approach of the active area of the sensitive detectors from the
beam. 
%While the acceptance for the 420+420 configuration in the 120 GeV range is 
%not too sensitive to the distance of approach, the acceptance of the 220+420 
%configuration is quite sensitive \cite{FP420TDR}. This is because the 220~m 
%detectors have acceptance only for relatively high $\xi$ forcing the proton 
%detected at 420~m to have lower $\xi$ and therefore to be closer to the beam. 
The final distance of approach
will depend on the beam conditions, machine-induced backgrounds and collimator
positions, and the RF impact of the detector on the beams. At 420~m the
nominal operating position is assumed to be between 5 and 7.5~mm, at 220~m it 
is between 2.0 and 2.5~mm. For masses above
about 120~GeV, the 220~m detector adds to the acceptance with increasing
importance as the central mass increases. The differences between ATLAS and CMS
acceptances for the 420+420 as well as 420+220 configurations (see 
Fig.~\ref{mh-acc}) are due to a different crossing angle which is in the 
vertical plane for IP1 (ATLAS) and horizontal plane at IP5 (CMS).
\vspace*{-0.3cm}
\subsection{Timing detectors}\label{timing}
The most prominent background to many diffractive physics analyses comes from 
an overlap of two soft SD events from pile-up and one ND event produced at a 
hard scale.
%The necessity of equipping the forward detectors operating at the highest
%luminosities by timing detectors emerged from studies estimating the effect
%of pile-up background on diffractive processes at LHC \cite{CMS-Totem}.
%At ${\cal L}=2\cdot 10^{33}$cm$^{-2}$s$^{-1}$, the
%average
%number is 7 events per bunch crossing, at $1\cdot 10^{34}$cm$^{-2}$s$^{-1}$,
%it is 35. Of these pile-up events, 3\% (1\%) contain a proton within the
%acceptance of forward detector at 220 m (420 m). If we consider e.g. the case
%of CEP of Higgs boson with mass of 120~GeV  that decays
%into a pair of b-jets, an overlay of three events, namely two SD events
%each with a proton within the acceptance of forward detectors on
%opposite sides and one hard-scale dijet event, mimics the signal almost
%perfectly. Given the much larger cross section of inclusive dijet events
%compared to the signal, this is the most important source of background. This
%background can be reduced by exploiting the correlations between quantities
%measured in the central detector and those measured by the forward detectors.
%One possibility is to compare $\xi$
%or pseudorapidity, $\eta$, another possibility is to use fast timing detectors
%placed close to the forward detectors. 
Fast timing detectors with an
expected sub-10 ps time resolution corresponding to a vertex resolution of
better than 2.1 mm should be able to assign a
vertex to the proton detected in the FPD and to reject about 97\% 
of cases that appear to be CEP events
but where the protons in reality originated from coincidences with pile-up
events. 
Presently two detector options are studied, namely
Quartz and Gas Cerenkov which may be read out with a Constant Fraction
Discriminator allowing the time resolution to be significantly improved
compared to usual electronics.

\subsection{Trigger}\label{trigger}
Due to a limited L1 trigger latency, detectors at 420~m are far away from the 
central detectors to be included in the L1 trigger in normal running 
conditions, while detectors at 220~m can in principle be included. 
The trigger strategy depends on the mass of the diffractively produced object
\cite{CMS-Totem}.
Demanding standard L1 triggers such as
those for high mass H$\rightarrow$WW/ZZ or high-$p_T$
dijet trigger would result in an acceptable output rate which may be further 
reduced by requiring the double proton tag at 220~m.
%The expected L1 rate of a dijet trigger with $E_T>100$ GeV is about 6~kHz at 
%the highest luminosity and it can be reduced to about 2~kHz if the double 
%proton tag at 220~m is required.

Triggering on low mass objects is more difficult but in principle feasible
as documented in \cite{Monika-trigger, AFPtrig} where diffractive
L1 triggers for a case of \hbb\ at $m_H$=120~GeV have been 
proposed. If the FPD trigger at 220~m is capable of triggering 
only on hits in the inner 4~mm part and if the L1 calorimeter is capable of 
defining exclusivity criteria using $E_T, \eta$ and $\phi$, then the final 
output rate is well below a 2~kHz limit at ${\cal L} = 
2\cdot 10^{33}$cm$^{-2}$s$^{-1}$ and slightly above this limit at ${\cal L} = 
10^{34}$cm$^{-2}$s$^{-1}$. Other reductions are under study \cite{AFPtrig}.

%Other reductions may be achieved from the 
%knowledge of the precise Higgs boson mass after it has been measured in other 
%processes and restrict accordingly (i) the dijet mass that could be calculated
%in the global merger processor or (ii) the average pseudorapidity that can be
%obtained using a finer segmentation of the L1 trigger at 220~m (0.5--1~mm 
%strips). This option is a part of the proposal for the trigger at 220~m based 
%on the quartz fibers \cite{220trigger}. \\

%\begin{eqnarray}
%c&=&|d|+|e|\nonumber\\
%&\stackrel{\text{(a)}}{=}&d+e\nonumber\\
%&\stackrel{\text{(b)}}{\geq}&\sqrt{f}\enspace,
%\end{eqnarray}
%\noindent where the equality (a) results from the fact that both
%$d$ and $e$ are positive while (b) comes from the definition of
%$f$.

\section{Acknowledgements}
Supported by the project AV0-Z10100502 of the Academy of Sciences of the Czech
republic and project LC527 of the Ministry of Education of the Czech republic.

%\section{Bibliography}
% ****************************************************************************
% BIBLIOGRAPHY AREA
% ****************************************************************************
\begin{footnotesize}
% IF YOU DO NOT USE BIBTEX, USE THE FOLLOWING SAMPLE SCHEME FOR THE REFERENCES
% ----------------------------------------------------------------------------

% ----------------------------------------------------------------------------

% IF YOU USE BIBTEX,
% - DELETE THE TEXT BETWEEN THE TWO ABOVE DASHED LINES
% - UNCOMMENT THE NEXT TWO LINES AND REPLACE 'Name_Of_Your_BibFile'

%\bibliographystyle{unsrt}
%\bibliography{Name_Of_Your_BibFile}
% example of Name_Of_Your_BibFile.bib
% @Article{Turcato:2006ch,
%      author    = "Turcato, M.",
%  collaboration = "ZEUS and H1",
%      title     = "Lepton flavour violation and charmonium physics at HERA",
%      journal   = "Nucl. Phys. Proc. Suppl.",
%      volume    = "162",
%      year      = "2006", 
%      pages     = "283-287",
%      SLACcitation  = "%%CITATION = NUPHZ,162,283;%%"
% }
% 
% @Unpublished{Gogitidze:2007du,
%      author    = "Gogitidze, N.",
%  collaboration = "H1", 
%      title     = "Prompt photons and particle momentum distributions at
%                   HERA", 
%      year      = "2007",
%      note    = "hep-ex/0701033",
%      SLACcitation  = "%%CITATION = HEP-EX 0701033;%%"
% }

\end{footnotesize}

% ****************************************************************************
% END OF BIBLIOGRAPHY AREA
% ****************************************************************************

\end{document}